# THE ENERGY SPECTRUM OF A RADIO PULSE CREATED BY A CASCADE DISK WITHIN THE MOON'S REGOLITH

Filonenko A.D. (e-mail: uy5lo@mail.ru), Nikitin E.V.  (e-mail: lovefamily@mail.ru)

East-Ukrainian National University,

kv. Molodiezhniy, 20A, Lugansk, Ukraine

ABSTRACT

Some known analytical solutions for a cascade shower were used for determining the intensity of excess electrons' radio-emission. It is shown that the energy spectrum has a sharp decline in the region of high frequencies. At a usual root-mean-square radius of the disk within the Moon's regolith ~ 0.06m the maximum of the spectrum is in the region ~ 0.5–0.6 GHz  while in the range 2–2.5 GHz the pulse's amplitude drops by 5-6 orders.  This fact is a consequence of loss of coherence due to finite dimensions of the cascade disk and demands conducting a significant correction of "radio-frequency window" for measuring cosmic rays' flux with a radio telescope.

It is known that motion of a charge with super-light velocity in a dielectric is accompanied by Cherenkov radiation.  In an idealized model of Cherenkov radiation at motion of a point charge with constant velocity through a homogeneous matter the duration of emitted radio pulse is indefinitely small and the radiation spectrum is practically unlimited if the angle of observation is   $\theta = \arccos(1/n)$. A real system of charges has finite dimensions, a limited region of motion and an observation angle different than the Cherenkov one.   In this paper, these emission parameters are investigated for a real object, which appears during the formation of a cascade shower of ultrahigh-energy particles ($10^{16} \div 10^{22}$ eV) in a gaseous or solid media. As a result of its interaction with the environment a clot of electrons and positrons, having in the maximum stage of the shower values quantities of the order of $10^7 \div 10^{13}$, is being formed. This cluster has the shape of a disk and moves faster than the speed of light in this environment. The thickness of the disk is thousands of times smaller than its radius. The highest charge density is achieved in a typical disk at its center, and one may approximately assume that the particle distribution has axial symmetry. Due to various processes associated with interaction of atoms of the medium with the particles of the shower, an excess of electrons in the disk is created whose energy corresponds to the Lorentz factor $\gamma \gg 1$ [1]. Therefore, the motion of particles of the disk, even in such a rarefied medium as the Earth's atmosphere, is accompanied by a coherent Cherenkov radiation in radio frequencies, because in standard atmosphere, with the rate $\gamma \sim 50 \div 100$ its velocity is greater than the velocity of electromagnetic wave [2].

Knowledge of the characteristics of the radiation field of the cascade disk is of interest in connection with the attempts to use coherent radiation for the detection of cosmic rays of ultrahigh energies. In this regard, since 1965, experiments were conducted studying the radio emission produced by extensive air showers (see, for example, the survey [3]). In [4] for the first time appeared the idea of the possibility of recording a radio pulse, caused by ultrahigh-energy particles on the Moon's surface using a radio telescope. Currently there are about a dozen experimental studies, but none of them have reliably reported signs of radio pulses from lunar origin [5 − 12].

A detailed study of the characteristics of the radiation with the help of software systems has been carried out in [13], whose results so far are essential for conducting experimental work. Parametrization of the field at the Cherenkov angle of observation is presented in this paper by the expression

$$| \vec{E}(\omega, R, \theta_C) | \sim \frac{\nu}{\nu_0} \frac{1}{1 + 0.4(\nu / \nu_0)^2} , \qquad (1)$$

where $\nu_0 = 0.5\,\text{GHz}$. The energy power spectrum of this radio pulse in relative units is shown in Fig.5 (curve 1).

Subsequently, these results were slightly modernized, but the main feature has remained virtually unchanged. It's the increase in the energy power spectrum $I(\omega) \sim |\vec{E}(\omega, R, \theta_C)|^2$ up to ~10–100 GHz [7,12] without a significant decrease. And consequently in later experiments the frequency range for the registration of radio pulses was chosen to be in the region of several GHz or even higher [10].

**On the pulse duration.**

Model of a shower in the form of a disk propagating in a homogeneous medium leads to some conclusions about the duration of the radio pulse. On Fig.1 AB is a disk with an excess of charged particles moving in the direction of $\vec{v}$. AD and BC − direction of radio waves incident on the antenna CD.

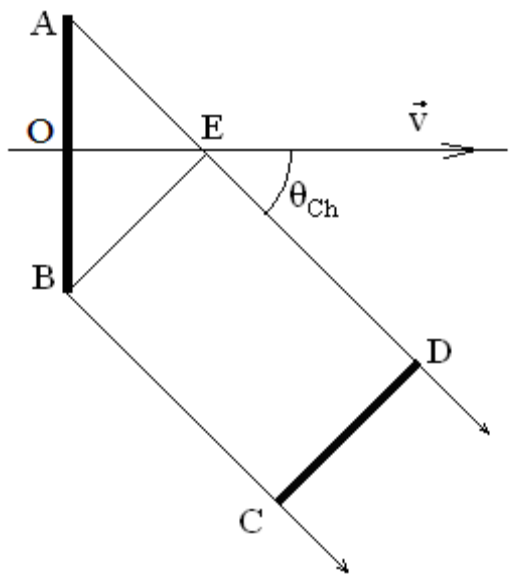

Fig.1.  Emission of the cascade disk at Cherenkov angle

OA=OB − rms radius of the disk.  Since its thickness is much smaller than the wavelengths considered here, we can assume that each point of AB emits coherently, and in the direction of AD and BC the phase difference of the extreme points of the disk is determined by the time τ of wave propagation in the area of AE. On this basis, we can write $\tau \approx (AB / v)n \sin \theta_{Ch}$, where n=1,7 − index of refraction and v~c − the speed of light in the lunar regolith. For typical sizes of AB ~ 12cm we ge length close to the value $\tau = 5.5 \times 10^{-10}$ s = 0.55 ns. This duration differs significantly from the results of calculations [13]. This refers to the following.  Expression (1) allows building a radio pulse as a function of time using the Fourier transform. The phase required for this is presented in Fig. 16 [13]. According to [13] in the frequency range 0 − 10 GHz the phase varies by no more than 40 degrees. It means that the phase dependence on frequency can be almost neglected, and we can write the inverse Fourier transform in the form of

$$E(t) \sim \int_{-\omega_l}^{\omega_l} \frac{\nu}{\nu_0} \frac{1}{1 + 0.4(\nu / \nu_0)^2} e^{i\omega t} d\omega \ . \tag{2}$$

The result of this recovery is shown in Fig.2 (curve 1).

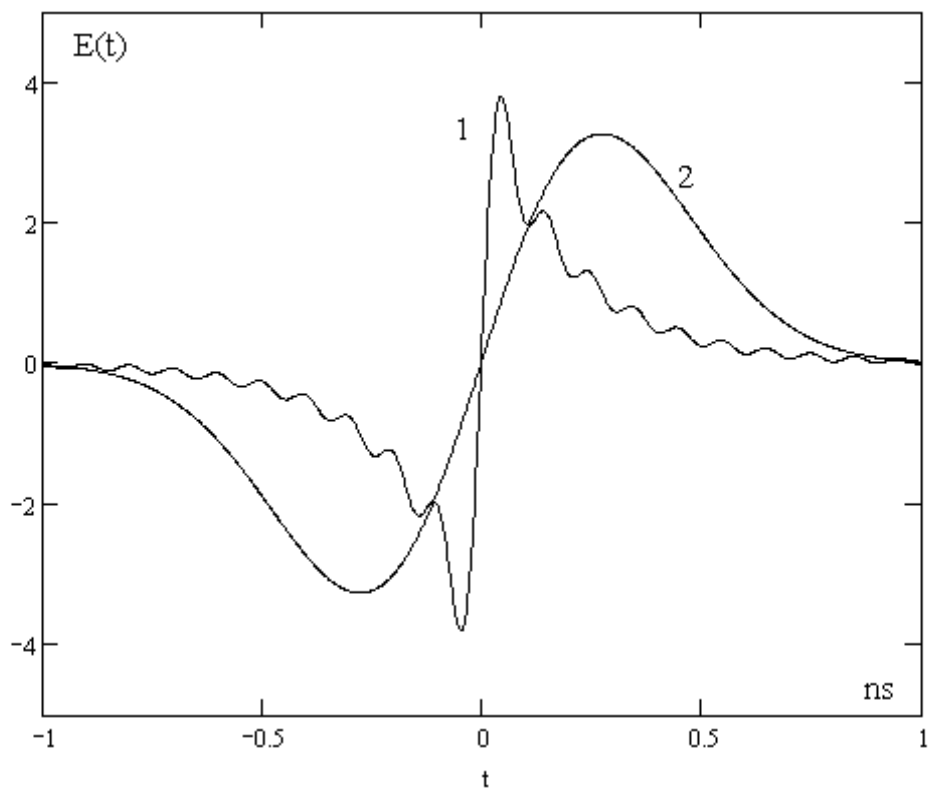

Fig.2. Pulse of the field strength, obtained from expression (1) -
curve 1, and the intensity corresponding to the spectrum (13) -  curve 2.
The scales on the axis $E(\nu)$  arbitrary

The time of the field strength step at t = 0 reaches infinitely small values $(<10^{-1} \, \text{ns})$ when $\omega_l \to 10 \, \text{GHz}$. This fact could explain the growth of the spectrum to such high frequencies (~ 10 GHz). However, above this frequency, the phase is unknown and so this explanation is not convincing. However, the expression (2) for the upper limit $\omega_l = 10 \, \text{GHz}$ shows that the time of the jump is 0.08 ns. This data, together with the expected duration $\tau = 5.5 \times 10^{-10}\,\text{s} = 0.55\,\text{ns}$ is very alarming and stimulate to review the expression (1) and modernize it, as it's performed in several works.

For example, in the experiment GLUE [7] they used the parametrization of the spectral energy in the form of a radio pulse (Fig.3, curve 1)

$$I(\omega, R, \theta_C)\,|\sim \left( \frac{\nu}{\nu_0} \frac{1}{1 + 0.4(\nu/\nu_0)^{1.44}} \right)^2, \tag{3}$$

where $\nu_0 = 2.5\,\text{GHz}$. In the NuMoon project [11] they used parametrization (Fig.3, curve 2)

$$I(\omega, R, \theta_C)\,|\sim \left( \frac{\nu}{\nu_0} \frac{1}{1 + (\nu/\nu_0)^{1.44}} \right)^2, \tag{4}$$

where $\nu_0 = 2.5\,\text{GHz}$ and in the RESUN project (12) they calculated the intensity according to expression (Fig.3, curve 3)

$$I(\omega, R, \theta_C)\,|\sim \left( \frac{\nu}{\nu_0} \frac{1}{1 + (\nu/\nu_0)^{1.23}} \right)^2, \tag{5}$$

where $\nu_0 = 2.23\,\text{GHz}$. The spectra (3-5), as shown in Fig.3, remained practically unabated up to 100 GHz, and this fact raises serious doubts. There is no physically based reason for this strange behavior of spectral functions.

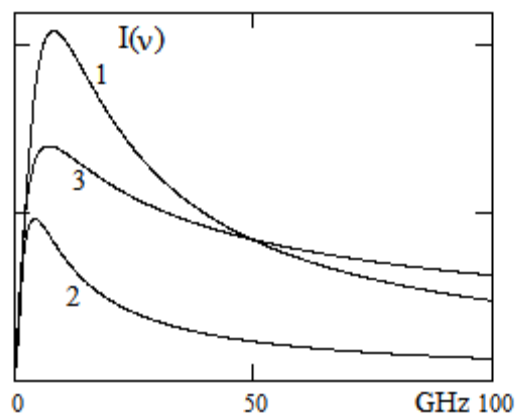

Fig. 3. Spectra $1 - 3$, respectively, for the papers [7,11,12]
shown in arbitrary scale on the axis $I(\nu)$

**The model of the cascade shower in the form of a current, modulated in amplitude**

The radiation field of the whole shower is the sum of the fields of the excessive electrons. The modern computing systems allow us to count the number of excess electrons at any stage of the shower, their energy distribution, spatial distribution and other parameters. To correct the addition of fields generated by each electron, one must know the phase of the field at the observation point. Depending on how well the dependence of the phase of the electromagnetic field on the above parameters was found, one can get all sorts of energy spectra of the radio pulse, including parameterization (3-5), too.

The spectrum of an electron moving with constant velocity on a limited interval L is a good example of the source of the radio pulse, whose spectrum is practically unlimited. We write the current density, generated by single electron moving uniformly along the axis with speed $\vec{v}$, in the form of

$\vec{j}(\vec{r}', t) = e\vec{v}\delta(z'-vt)\delta(x')\delta(y')$, where $\vec{r}' = \vec{i}x' + \vec{j}y' + \vec{k}z'$ is a radius-vector of some point. The Fourier component can be found using the expression

$$\vec{j}(\vec{r}', \omega) = q\vec{v}\int_{-\tau}^{\tau}\delta(z'-vt)\delta(x')\delta(y')e^{i\omega t}dt = q\vec{e}_z e^{i\frac{\omega}{v}z'}\delta(x')\delta(y').$$

Far from the source at a distance r the vector potential can be found as [14]

$$\vec{A}(\vec{r}, \omega) = \frac{\mu_0 e^{ikr}}{4\pi r}\int_{V'}\vec{j}(\vec{r}', \omega)e^{-i\cdot\vec{k}\vec{r}'}dV' = \frac{\mu_0 e^{ikr}}{4\pi r}q\vec{e}_z\frac{2\sin[kL(1-\beta n\cos\theta)]}{k(1-\beta n\cos\theta)}.$$

We finally obtain for the spectrum I(ω) of the radio-signal

$$I(\omega) \sim |\vec{B}(\omega)|^2 = |rotA(\omega)|^2 \sim \left(\frac{\sin[kL(1-\beta n\cos\theta)]}{(1-\beta n\cos\theta)}\right)^2, \tag{6}$$

where β=v/c is the relative velocity of the charge, n − index of refraction of the medium and θ − observation angle. The spectrum (6) decreases with increasing frequency during the observation at any angle. This signifies, in particular, that in case of unsuccessful addition of fields of individual electrons one can obtain spectra similar to (3-5).

However, another approach to the problem of the addition of fields is possible. For example, calculating the radiation field from the conductor, we do not watch the phases of a single electron, it is sufficient to know the function of current density at each point of the conductor. The motion of a charged disk of the cascade shower in this method is a current, too. The radiation field in this case can be found if one knows the cascade function. Some help can be obtained from using the information on the angular distribution of electrons in relation to the axis of the shower, since the sum of the transverse components of current densities of excess electrons does not contribute to the radiation field. However, the use of angular distribution seems to be important only at later stages of a shower which does not contribute substantially to current density function.

In line with this we express the current density caused by the motion of the cascade disk and obtain the radiation field according to the laws of classical electrodynamics. The cascade function in this case will act as the modulator of current. To make clear an example of this and avoid any connection with software of computer systems we define the cascade function in the form of approximation of analytical solution [15].

$$N(x) = \frac{0.31}{\sqrt{\alpha}} e^{\alpha - \frac{(z' - \alpha z_0)^2}{D \alpha z_0^2}}, \tag{7}$$

where $z' = vt$ - the depth of the shower in meters, $\alpha = \ln W_0 / W_{cr}$, $W_0$ − total energy of the shower, $W_{cr}$ − critical energy of electrons in the lunar regolith, $z_0 \approx 0.14\,m$ radiation length in meters.

Here are a few observations concerning the validity of the use of expression (7). It is known that to calculate the cascade functions one generally uses modern software systems, which take into account the LPM effect for dense media at high initial energies. Using the full set of the results in this paper for the problem is hardly advisable, primarily for reasons related to the use of unnecessarily large computational resources. In addition, an analytical solution retains the ability to verify the results of virtually every step of the solution. Therefore it's more  useful to parameterize the numerical calculations in the form of simple expressions, correctly displaying numerical solutions in a certain range of values of the variables.

To solve this problem an approximation of the cascade function must be made in such a way that it would retain the properties of the numerical solution, on which the basic parameters of the radio pulse depend. In this case it's, at least, the radiation pattern width at half maximum (ie, angular "thickness" of the Cherenkov cone) and the energy spectrum of the radio pulse. These parameters, in turn, are linked with an effective length of the cascade shower L (that is the path traversed by the disk at the ends of which the number of particles in the shower is 2-3 times less than at the maximum) and with a characteristic transverse size of the shower on this section of the route. To some extent (but not a decisive one) the spectrum is dependant on the shape of the cascade function within the interval L.

For example, we know from classical electrodynamics that the width of the radiation pattern of one-dimensional current system directly depends on its size. Likewise, the length of the small charge's trajectory form the diagram of Cerenkov radiation. The size of a charged body almost completely determine the radiation spectrum.

In this paper, we use as cascade function expression (7), which is a simplified version of the approximation [15].

$$N(W_0, \chi) = \frac{0.31}{\sqrt{\alpha}} e^{\chi\left(1 - \frac{3}{2}\ln\frac{3\chi}{\chi + 2\alpha + 2x}\right)},$$

where $\chi$ − the depth of the shower in avalanche units. In expression (1) we take into account that the quantity $x = \ln(r/r_M)$ is small in comparison with $\chi + 2\alpha$. Factor D does not affect the maximum number of particles in the shower, but an effective path of a cascade shower is totally dependent on it. According to the parameterization of numerical calculation of one-dimensional cascade function, the effective length of a cascade shower in the lunar regolith is equal to [16]

$$L(x) = 12.7 + \frac{2}{3} \log_{10}\left(\frac{W_0}{10^{20} \text{ eV}}\right).$$

At energy $W_0 = 10^{22}$ eV the length is equal to L=14 avalanche regolith units or $L = 14 \times (X_0 / \rho) = 1.82$ m for $\rho = 1.7 \text{ g/cm}^3$ and $X_0 = 22.1 \text{ g/cm}^2$. For such value of L factor D in expression (1) corresponds to 2. Radiation unit of length $X_0 = 22.1 \text{ g/cm}^2$ and regolith volume density $\rho = 1.7 \text{ g/cm}^3$ are not exactly determined values and depend on local conditions at the Moon's surface. That's why real D may be much different than D=2. Besides, one must add that we may put an arbitrary value for D and obtain results for any effective length of the shower.

Radial distribution of particle density of the cascade disk in relation to the cascade shower axis is random. In order to make the model closer to real characteristics one can choose normal distribution law over radius r, the fact that in its main features is confirmed by experimental measurements in extensive air showers

$$\rho(r) = \rho_0 e^{-\frac{r^2}{2r_l^2}} \tag{8}$$

The value $r_l$ plays the role of the root-mean-square radius and is determined by the condition in which a cylinder of length 2b and radius $r_m$ contains half of the particles.

$$q = \rho_0 2b \int_V e^{-\frac{r^2}{2r_l^2}} r dr d\varphi dz = 4\pi \rho_0 b \int_0^\infty e^{-\frac{r^2}{2r_l^2}} r dr = 4\pi \rho_0 b r_l^2 \int_0^\infty e^{-\frac{r^2}{2r_l^2}} d\left(\frac{r^2}{2r_l^2}\right) = 4\pi \rho_0 b r_l^2$$

It means that:

$$2\pi \rho_0 b r_l^2 = 4\pi \rho_0 b r_l^2 \left(1 - e^{-r_m^2/2r_l^2}\right)$$ or $r_m / r_l = 1.177$, i.e. $r_l$ – is the radius of the circle that contains about half of the particles of the disc. Let us divide the charged body having cylindrical symmetry into elementary volumes dV'', each of which moves along the Z axis with the same velocity $\vec{e}_z v$ (v ~ c) and all the way through the magnitude of each of the elementary charges of the system $\rho(x'', y'', z'') dV''$ is modulated in amplitude by the factor (7).

Inside the body we use frame of reference $(X'', Y'', Z'')$ and each of the elements $dV''$ has coordinates $(x'', y'', z'')$. Then the charge density created by the element $dV''$ at some point $(x', y', z')$ is equal to

$$\vec{j}(x', y', z', t) = \vec{e}_z v dV'' \rho_0 e^{-\frac{r''^2}{2r_1^2} - \frac{(vt - \alpha z_0)^2}{3\alpha z_0^2}} \delta[(z' - z'') - vt] \delta(y' - y'') \delta(x' - x''), \quad (9)$$

where $\vec{r}'' = \vec{i} x'' + \vec{j} y'' + \vec{k} z''$. We use delta function for describing the charge density. The expression (9) shows that only within an infinitesimal element $dV''$ at the time $t = (z' - z'') / v$ the current density is not zero.

The Fourier component of the charge density (9) is equal to

$$\vec{j}(r', \omega) = \vec{e}_z v dV'' \rho_0 e^{-\frac{r''^2}{2r_1^2}} \int_{-\infty}^{\infty} e^{i\omega t} e^{-\frac{(vt - \alpha z_0)^2}{3\alpha z_0^2}} \delta(x' - x'') \delta(y' - y'') \delta[(z' - z'') - vt] dt =$$

$$(10)$$

$$= \vec{e}_z dV'' \rho_0 e^{-\frac{r''^2}{2r_1^2}} e^{-\frac{[z' - (z'' - \alpha z_0)]^2}{3\alpha z_0^2}} e^{i\frac{\omega}{v}(z' - z'')} \delta(x' - x'') \delta(y' - y'')$$

The observer is at a distance much larger than the size of the system of charges, so the vector potential created by the element $\rho dV''$ in accordance with (10), can be written as [14]

$$d\vec{A}(\vec{r}, \omega) = \vec{e}_z dV'' \frac{\mu_0 e^{iknr}}{4\pi r} \rho_0 e^{-\frac{r''^2}{2r_1^2}} \int_{V'} e^{i\frac{\omega}{v}(z' - z'')} e^{-\frac{[z' - (z'' - \alpha z_0)]^2}{3\alpha z_0^2}} \delta(x' - x'') \delta(y' - y'') e^{-in\vec{k}\vec{r}'} dV' =$$

$$= \vec{e}_z dV'' \frac{\mu_0 e^{iknr}}{4\pi r} \rho_0 e^{-\frac{r''^2}{2r_1^2}} \int_{-\infty}^{\infty} e^{ik(z' - z'')} e^{-\frac{[z' - (z'' - \alpha z_0)]^2}{3\alpha z_0^2}} e^{-iknz'\cos\theta - i(k_x n x'' + k_y n y'')} dz'$$

$$(11)$$

where $dV' = dx' dy' dz'$ − stationary element with coordinates $(x', y', z')$, $r$ − the distance from the origin of $(X, Y, Z)$ to the observer. Let's rotate the unprimed coordinate system $(X, Y, Z)$ so that the wave propagation vector projection $\vec{k}$ on the plane $(X, Y)$ coincided with the X-axis, and the axes $(X'', Y'', Z'')$ coincided with $(X, Y, Z)$ axes. Then $k_y n y'' = 0$, $k_x = k \sin\theta$, and $x'' = r \cos\varphi$.

Let's write (11) as

$$d\vec{A}(\vec{r}, \omega) = \vec{e}_z dV'' \frac{\mu_0 e^{iknr}}{4\pi r} \rho_0 e^{-\frac{r''^2}{2r_1^2}} e^{-iknz''\cos\theta} e^{-iknr''\cos\varphi \sin\theta} e^{-ik\alpha z_0 p} \int_{-\infty}^{\infty} e^{ikup} e^{-\frac{u^2}{3\alpha z_0^2}} dz'$$

(12)

where p=(1−βncosθ) and β=v/c≈0.9998≈1 − relative velocity of the disk. The integral in this expression is equal to $z_0\sqrt{3\alpha\pi}\exp[-3\alpha k^2 p^2 z_0^2/4]$.

Let's find the vector potential of the whole shower in cylindrical coordinates (see Fig.4) as the sum over the whole elements of dV''= rdφdzdr .

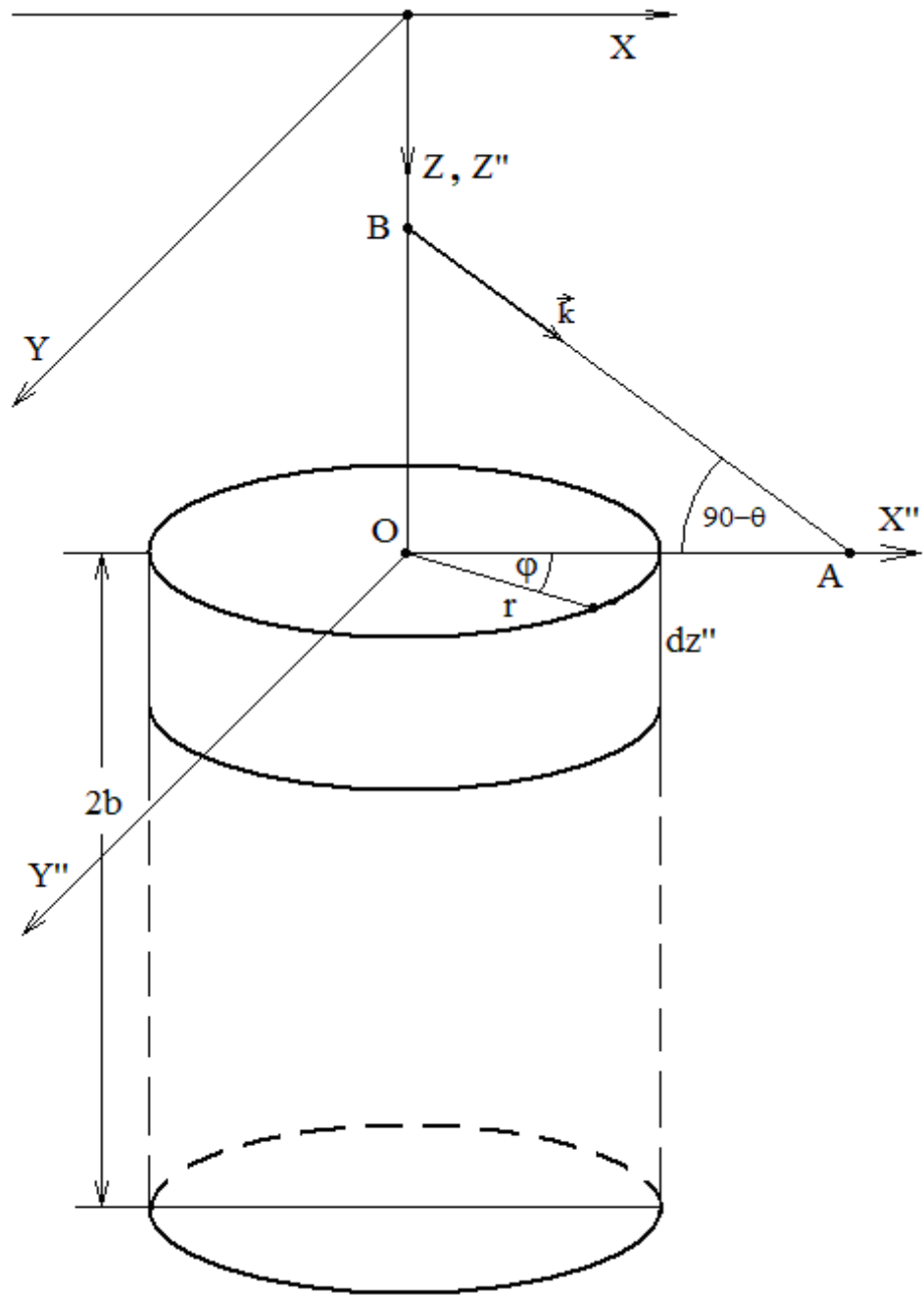

Fig. 4

Fig. 4. Details for calculation of the vector potential in cylindrical coordinates

$$\vec{A}(\vec{r},\omega)=\vec{e}_z\frac{\mu_0e^{iknr}}{4\pi r}\rho_0z_0\sqrt{3\alpha\pi}e^{-3\alpha k^2p^2z_0^2/4}e^{-ik\alpha z_0p}\int\limits_{-b}^{b}\int\limits_{0}^{2\pi}\int\limits_{0}^{\infty}e^{-\frac{r''^2}{2r_1^2}}e^{-iknr''\cos\varphi\cos\theta}r''dr''d\varphi dz''=$$

$$=\vec{e}_z\frac{\mu_0e^{iknr}}{4\pi r}\rho_0z_0\sqrt{3\alpha\pi}e^{-3\alpha k^2p^2z_0^2/4}e^{-ik\alpha z_0p}\frac{2\sin(knb\cos\theta)}{kn\cos\theta}\int\limits_{0}^{\infty}\int\limits_{0}^{2\pi}e^{-\frac{r''^2}{2r_1^2}}e^{-iknr''\cos\varphi\sin\theta}r''dr''d\varphi=$$

$$\vec{e}_z\frac{\mu_0e^{iknr}}{r}\rho_0z_0\sqrt{3\alpha\pi}e^{-3\alpha k^2p^2z_0^2/4}e^{-ik\alpha z_0p}\frac{\sin(knb\cos\theta)}{kn\cos\theta}\int\limits_{0}^{\infty}e^{-\frac{r''^2}{2r_1^2}}J_0(knr''\sin\theta)r''dr''$$

Finally we find the dependence of intensity of the signal on the frequency ($k=n\omega/c$) and the angle $I(k,\theta)\sim|\vec{B}|^2=|\operatorname{rot}\vec{A}|^2$

$$I(k,\theta)\sim\left|e^{-3\alpha k^2(1-n\cos\theta)^2z_0^2/4}\frac{\sin(knb\cos\theta)}{nctg\theta}\int\limits_{0}^{\infty}e^{-\frac{r''^2}{2r_1^2}}J_0(knr''\sin\theta)r''dr''\right|^2 \qquad (13)$$

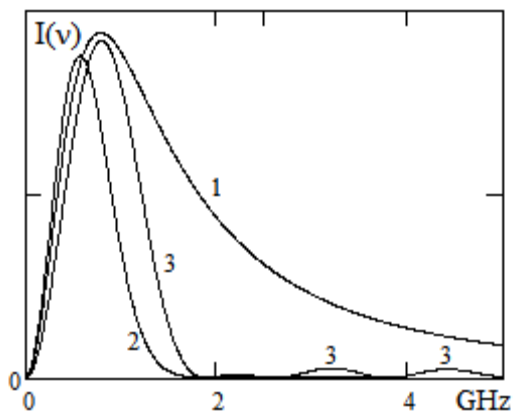

Fig. 5. The energy spectra of radiation [13] − curve 1, for the expression

(13) - curve 2 and for a cube-shaped body (14) - curve 3. The scales on

axis $I(\nu)$ arbitrary

In Fig. 5 (curve 2) in relative units the spectrum (13) is shown with typical values $\alpha=33$, $z_0=0.14\,\mathrm{m}$, $b=10^{-3}\,\mathrm{m}$, $r_1=0.06\,\mathrm{m}$, n=1.7, $\theta$=0.942 (the Cherenkov angle). The maximum is reached at a frequency of 0.56 GHz. At a frequency of $\sim 2.5$ GHz the intensity falls by nearly 6 orders of magnitude. Field E (t) as a function of time is shown in Fig.2 (curve 2). Pulse duration between the maximum and minimum is 0.55 ns. For curve 1 (according to expression (2)), this value is 0.08 ns.

It may seem that the spectrum shown in Fig. 5 (curve 2), is the result of selection of the cascade function or of a definite choice of the normal distribution over the particle radius (7-8). However, it is easy to see that even in the case when we choose a constant instead of the cascade function (this assumption can only enrich the spectrum with high frequencies) and assume that the density distribution over the volume of the body V" remains unchanged, we get a spectrum close to the one in (13). The curve 3 in Fig. 5 in relative terms is the spectrum of a homogeneously charged body, cube-shaped with side b=0.06 m, moving uniformly on the segment 2L=1.6 m in a matter with n=1.7.

$$I(k,\theta) \sim |\vec{B}|^2 = |rot\vec{A}|^2 \sim \left| \frac{\sin[k(1-n\cos\theta)L]}{k(1-n\cos\theta)} \frac{\sin(kb)}{k} \frac{\sin(nkb\sin\theta)}{nk\sin\theta} \right|^2. \qquad (14)$$

The maximum of this spectrum (14) is at a frequency of 0.8 GHz. At 2.5 GHz the intensity decreases by three orders of magnitude. The motion of a body in the form of a cube has nothing common with the one of the cascade shower except its super-light velocity, but the spectrum (14) is very close to (13). Therefore, a sharp decline in the closest region to the maximum field is rather a rule than a coincidence.

### References

1. G.A. Askar'yan. Soviet Phys. JETF, 14, 441, 1962

2. G.A. Askar'yan. Soviet Phys. JETF, 21, 658, (1965)

3. A.D. Filonenko.    Phys. Usp. **45** (4), 403–432 (2002)

4. R.D. Dagkesamansky, I.M. Zheleznyk. JETF 50, 233, (1989)

5. T.H. Hankins, R.D.  Ekers, D.O. O'Sullivan. Monthly Notices Roy. Astron. Soc., **283,** 1027, (1996)

6. P.W. Gorham, K. M. Liewer,  C. J. Naudet. arXiv:astro-ph/9906504

7. P. W. Gorham,  K. M. Liewer, C. J. Naudet. et al. arXiv:astro-ph/0102435

8. P. W. Gorham, C. L. Hebert,   K. M. Liewer. et al. arXiv astro-ph/0310232

9. E.P. Abranin, P.I. Golubnichij, A.D. Filonenko. Izvestiya Rossiiskoi Akademii Nauk. Seriya Fizicheskaya (Bulletin of the Russian Academy of Sciences: Physics) v.65, №11, p. 1670-1671, (2001)

10. A.R. Beresnyak1, R.D. Dagkesamanski, I.M. Zheleznykh. Astronomy Reports,Vol. 49, No. 2, pp.127–133, (2005)

11. S. Buitink, J. Bacelar, R. Braun. et al. arXiv: 0808.1878 [astro-ph]

12. T. R. Jaeger, Mutel R. L., Gayley K. G. arXiv 0910.5949 [astro-ph.IM]

13. E. Zas, F. Halzen, T. Stanev. Physical Review **D,** №1, v.45, p.362-376. (1992)

14. L. D. Landau and E. M. Lifshitz. The Classical Theory of Fields (Cambridge, Mass.: Addison-Wesley, 1951).

15. S. Hayakawa. Cosmic ray physics nuclear and astrophysical aspects. 1969

16. O. Scholten, J. Bacelar, R. Braun et al. arXiv:astro-ph/0508580